\documentclass[amssymb,prb,twocolumn,superscriptaddress,floats,showkeys,showpacs]{revtex4-1}

\usepackage{float}
\usepackage[caption=false]{subfig}
\usepackage{ulem}
\usepackage[euler]{textgreek}
\usepackage{mathrsfs}
\usepackage[T1]{fontenc}
\usepackage{bm}
\usepackage{graphicx}% Include figure files
\usepackage{natbib}
\usepackage{amsmath}
\usepackage{textcomp}
\usepackage{epstopdf}
\usepackage{hyperref} 
\hypersetup{colorlinks,citecolor=blue, filecolor=blue ,linkcolor=blue , urlcolor=blue, pdftex}
\usepackage{sidecap}

\def \FUW{Institute of Experimental Physics, Faculty of Physics, University of Warsaw, ul. Pasteura 5, 02-093 Warszawa, Poland}

\begin{document}

\title{Resonance and antiresonance in Raman scattering in GaSe and InSe crystals}

\author{M. Osiekowicz}\affiliation{\FUW}
\author{D. Staszczuk}\affiliation{\FUW}
\author{K. Olkowska-Pucko}\affiliation{\FUW}
\author{\L{}. Kipczak}\affiliation{\FUW}
\author{M. Grzeszczyk}\affiliation{\FUW}
\author{M. Zinkiewicz}\affiliation{\FUW}
\author{K.~Nogajewski}\affiliation{\FUW}
\author{Z.~R. Kudrynskyi}\affiliation{School of Physics and Astronomy, The University of Nottingham, Nottingham, NG7 2RD, UK}
\author{Z. D. Kovalyuk}\affiliation{Institute for Problems of Materials Science, The National Academy of Sciences of Ukraine, Chernivtsi, 58001, Ukraine}
\author{A. Patan\'e}\affiliation{School of Physics and Astronomy, The University of Nottingham, Nottingham, NG7 2RD, UK}
\author{A. Babi\'nski}\affiliation{\FUW}
\author{M. R. Molas}\email{maciej.molas@fuw.edu.pl}\affiliation{\FUW}

\begin{abstract}
	
The temperature effect on the Raman scattering efficiency is investigated in $\varepsilon$-GaSe and $\gamma$-InSe crystals. We found that varying the temperature over a broad range from 5~K to 350~K permits to achieve both the resonant conditions and the antiresonance behaviour in Raman scattering of the studied materials. The resonant conditions of Raman scattering are observed at about 270~K under the 1.96~eV excitation for GaSe due to the energy proximity of the optical band gap. In the case of InSe, the resonant Raman spectra are apparent at about 50~K and 270~K under correspondingly the 2.41~eV and 2.54~eV excitations as a result of the energy proximity of the \mbox{so-called} B transition. Interestingly, the observed resonances for both materials are followed by an antiresonance behaviour noticeable at higher temperatures than the detected resonances. The significant variations of phonon-modes intensities can be explained in terms of electron-phonon coupling and quantum interference of contributions from different points of the Brillouin zone.
\end{abstract}

%\keywords{single quantum dot; micro-luminescence; valence band mixing; neutral exciton; triexciton;}
%\pacs{78.67.Hc, 71.35.-y, 78.55.Cr}
\maketitle

%XXXXXXXXXXXXXXXXXXXXXXXX        INTRO
\section{Introduction \label{sec:Intro}}

Two-dimensional (2D) van der Waals crystals have recently attracted considerable attention due to their unique electronic band structure and functionalities~\cite{novoselov2005,Geim2013}. The main focus of researchers has been on semiconducting transition metal dichalcogenides (\mbox{S-TMDs}), $e.g.$ MoS$_2$, WSe$_2$, and MoTe$_2$~\cite{koperski, Wang2018}. Currently, another much larger group of layered materials, $i.e.$ semiconducting post-transition metal chalcogenides (S-PTMCs), $e.g.$ SnS, GaS, InSe, and GaTe, has drawn the attention of the 2D community. Among these crystals, Se-based compounds of S-PTMCs, $i.e.$ InSe and GaSe, demonstrate a tunability of their optical response from the near infrared to the visible spectrum with decreasing layer thickness down to monolayers~\cite{Mudd2016,Bandurin2016,Terry2018}.

Raman scattering (RS) spectroscopy is a powerful and nondestructive tool to get useful information about material properties~\cite{Zhang2016}. The RS measurements provide an insight into their vibrational and electronic structures and are of particular importance in studies of layered materials ~\cite{Zhang2015a}. The flake thickness, strain, stability, charge transfer, stoichiometry, and stacking orders of the layers can be accessed by monitoring parameters of the observed phonon modes~\cite{Ferrari2006,Lee2010,Cancado2011,Tonndorf2013,Bruna2014,golasaAIP,Grzeszczyk2016,Kipczak2020}.
RS experiments can be performed under non-resonant and resonant excitation conditions:~\cite{Smith2004}. The resonant excitation may lead to a significant enhancement of the RS intensity in S-TMD as well as the activation of otherwise inactive modes. This offers supplementary information on the coupling of particular phonons to electronic transitions of a specific symmetry~\cite{Corro2014,molasSR,Shree2018}. The crossover between the non-resonant and resonant conditions can be achieved not only by the variation of the excitation energy but also by the modulation of temperature as it was recently reported ~\cite{GolasaNano,Grzeszczyk2018,Molas2019}. In such an approach, it is the band structure that changes with temperature allowing for resonance with particular excitation energy. 

In this work, we present a comprehensive investigation of the effect of temperature on the Raman scattering in $\varepsilon$-GaSe and $\gamma$-InSe crystals. It has been found that the intensity of some phonon modes exhibits a strong variation as a function of temperature under excitation with specific energy due to the resonant conditions of RS. Moreover, a significant antiresonance behaviour accompanies the resonances at higher temperatures, which leads to the vanishing of the modes intensities. The observed effects are discussed in terms of electron-phonon coupling and quantum interference of contributions from different points of the Brillouin zone (BZ).

\section{Samples and experimental setups \label{experiment}}

\begin{figure*}[t]
	\includegraphics[width=1\textwidth]{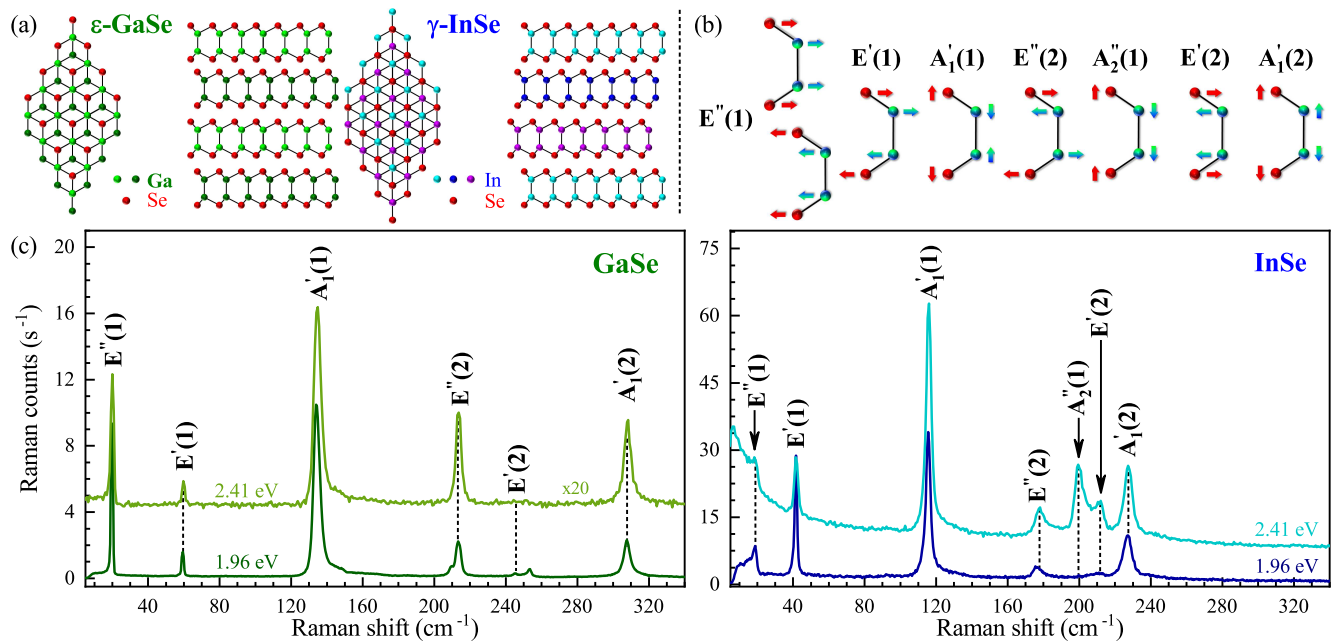}
	\caption{(a) Atomic structure and stacking order of $\varepsilon$-GaSe and $\gamma$-InSe. (b) Atomic displacements of the studied phonon modes. (c) unpolarized Raman scattering spectra of the GaSe (green/bright green) and InSe (blue/bright blue) crystals measured under two different excitation energies at room temperature. The spectra are shifted for clarity.} \label{fig:1}
\end{figure*}

The $\gamma$-InSe crystals were grown using the Bridgman method from a polycrystalline melt of In$_{1.03}$Se$_{0.97}$, while the $\varepsilon$-GaSe crystals were purchased from 2D Semiconductors. To facilitate carrying out the optical experiments, the investigated crystals were glued on Si/SiO$_2$ substrates.

The photoluminescence (PL) and RS measurements were performed using \mbox{$\lambda$=488~nm} (2.54~eV) and $\lambda$=514.5~nm (2.41~eV) ($\lambda$=632.8~nm (1.96~eV)) radiation from a continuous wave Ar-ion (He-Ne) laser. The studied samples were placed on a cold finger in a continuous flow cryostat mounted on $x$-$y$ manual positioners. The excitation light was focused by means of a 100x long-working distance objective with a 0.55 numerical aperture producing a spot of about 1 $\mu$m diameter. The signal was collected via the same microscope objective, sent through 1 m monochromator, and then detected by using a liquid nitrogen cooled charge-coupled device camera. The excitation power focused on the sample was kept at 200~$\mu$W during all measurements to avoid local heating. To detect low-energy Raman scattering up to about $\pm$10~cm$^{-1}$ from laser line, a set of Bragg filters was implemented in both excitation and detection paths. Moreover, the samples were mounted so that the 2D plane of crystals was at an angle of $\theta$=60$^{\circ}$ relative to the excitation/detection light direction to avoid the Rayleigh scattering during measurements of RS spectra.

\section{Experimental results \label{raman}}

GaSe or InSe crystals are layered materials composed of monolayers stacked by weak van der Waals forces. A monolayer of GaSe or InSe crystals comprises four covalently bonded atomic planes (Se-Ga-Ga-Se or Se-In-In-Se) arranged in a hexagonal atomic lattice~\cite{Schubert1954,Semiletov1958}. Bulk materials or thicker layers, achieved by stacking of monolayers by weak van der Waals forces, can exist in multiple stacking orders (polytypes)~\cite{Jandl1978}. The GaSe crystals are most commonly found in $\varepsilon$ polytype with the unit cell having 8 atoms and spanning two layers. The majority of the InSe crystals stacks in the $\gamma$ polytype, in which the unit cell extends over three structural layers and contains 12 atoms. The crystal structures of the studied $\varepsilon$-GaSe and $\gamma$-InSe thin layers are presented schematically in Fig.~\ref{fig:1}(a). Note that the polytype indications are omitted in the following.

\subsection{Effect of the excitation energy on the Raman scattering \label{results:resonant}}

Atomic displacements of the phonon modes apparent in investigated GaSe and InSe crystals are shown in Fig.~\ref{fig:1}(b). The corresponding unpolarized RS spectra of GaSe and InSe obtained under two different excitation energies (1.96~eV and 2.41~eV) measured at room temperature ($T$=300~K) are presented in Fig.~\ref{fig:1}(c). The RS peaks are classified according to their irreducible representations in the symmetry group in the monolayer phase D$_\textrm{3h}$ and additionally numbered due to their increased Raman shift for clarity. Note that the detailed description of the optical modes active in RS and infrared absorption can be found in previous works devoted to the RS spectroscopy performed on the GaSe~\cite{Hoff1975, Gasanly2002} and InSe~\cite{Jandl1978,Kuroda1978}. As can be appreciated in Fig.~\ref{fig:1}(c), the number of observed phonon peaks and their intensities are substantially affected by the excitation energy. Particularly, all RS spectra exhibit three in-plane phonon modes (E$^{''}$(1), E$^{'}$(1), and E$^{''}$(2)) and two out-of-plane phonon modes (A$_1^{'}$(1) and A$_1^{'}$(2)), which are observed regardless of the excitation energy. For GaSe, a significant increase in the RS intensity of about 20 times is observed while the sample is excited with 1.96~eV as compared to 2.41~eV. In the case of InSe, the difference in the RS intensities obtained for both excitation energies can be neglected, while a broad emission extending from laser liner towards higher Raman shifts is observed under excitation of 2.41~eV. There are also additional Raman peaks under excitation of 1.96~eV and 2.41~eV for GaSe and InSe, respectively. The most commonly reported, such as out-of-plane A$_1^{''}$(1) and in-plane E$^{'}$(2), are labelled in the RS spectra. Since GaSe is a polar crystal~\cite{Wieting1972}, the assignment of other peaks apparent in the RS spectrum of GaSe excited with 1.96 eV is more complex (see Ref.~\citenum{Irwin1973,Hoff1974,Hoff1975} for details).

\begin{figure*}[t]
	\includegraphics[width=1\textwidth]{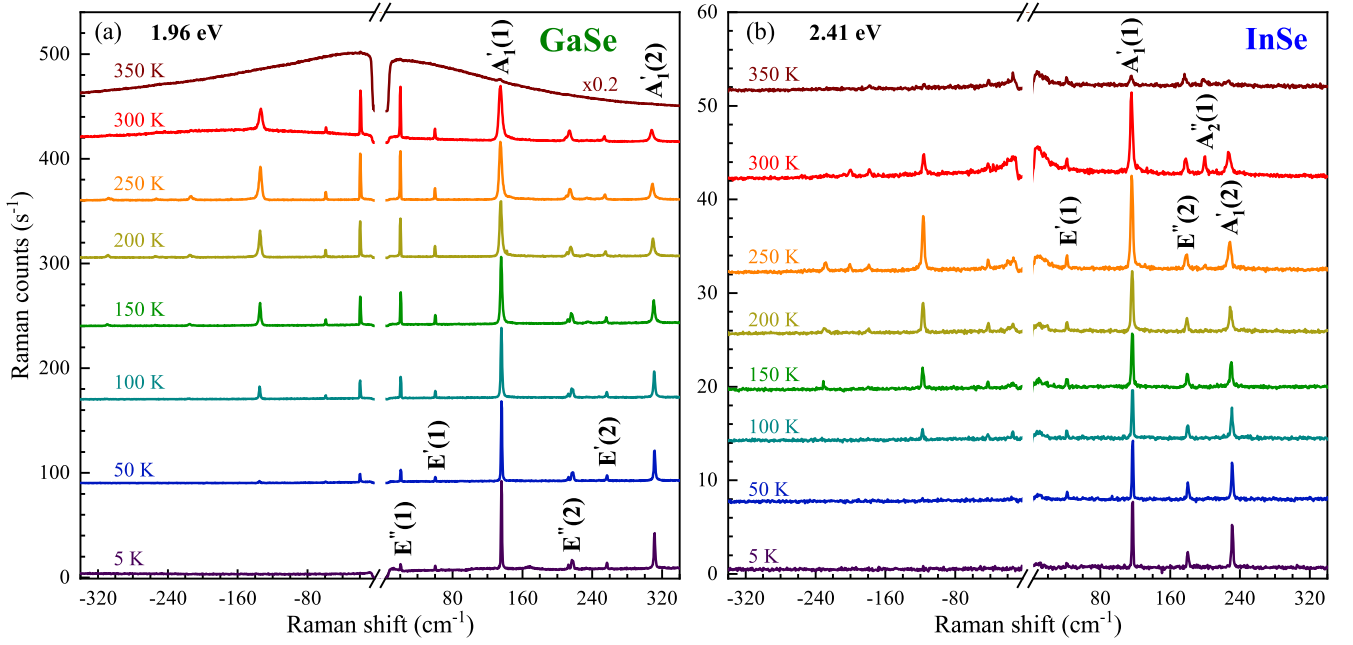}
	\caption{Raman scattering spectra of the studied (a) GaSe and (b) InSe measured at selected temperatures under resonant excitation regimes. The spectra are shifted for clarity.} \label{fig:2}
\end{figure*}

To understand the observed difference in the shapes and the intensities of the measured RS spectra of GaSe and InSe crystals, their band structures need to be discussed. The optical band-gap energy of the GaSe at room temperature was reported to be of around 2.0~eV~\cite{Terry2018}, while the corresponding optical gap energy of the InSe is found to be on the order of 1.25~eV~\cite{Mudd2016,Bandurin2016}. In consequence, the significant increase of RS intensity in GaSe under excitation of 1.96~eV can be related to the proximity of the optical band gap. The obtained difference in RS in InSe is more striking as compared to the GaSe case, due to both excitation energies (1.96~eV and 2.41~eV) being much larger than the InSe optical band gap. However, the 2.41~eV excitation stays in the vicinity of the so-called B transition at $\Gamma$ point of the BZ in InSe of energy around 2.4~eV~\cite{Bandurin2016}. It suggests that resonant conditions of Raman scattering can be achieved under excitation of 1.96~eV for GaSe and 2.41~eV for InSe due to the energy proximity of the optical band gap and the B transition, respectively. 

%To study the effect of temperature on the Raman spectra under the resonant excitation regimes, 

\subsection{Temperature influence on resonant Raman scattering \label{results:temperature}}

In order to investigate the aforementioned resonant conditions in Raman scattering, we carried out RS experiments in a broad range of temperatures from about 5~K to 350~K. Figs~\ref{fig:2}(a) and \ref{fig:2}(b) display the RS spectra of the GaSe and InSe consisting of both the Stokes and anti-Stokes scatterings measured at selected values of temperature under excitations of 1.96~eV and 2.41~eV, respectively. As can be appreciated in the Figure, only the Stokes-related signal is observed at the lowest temperatures. It is due to the nature of Stokes and anti-Stokes processes, which are related correspondingly to emission and absorption of phonons. As phonons are bosons, their population is described by the Bose-Einstein statistics. At absolute zero temperature, a crystal lattice lies in its ground state and contains no phonons. As a result, an increase of sample temperature leads to increased number of available phonons, which is accompanied by an increase of the number of phonon absorption processes giving rise to the anti-Stokes signal at elevated temperatures. This kind of evolution is demonstrated in Figs~\ref{fig:2}(a) and \ref{fig:2}(b). Moreover, the influence of increasing temperature on a given RS peak is revealed by a red shift of energy, a broadening of linewidth, and a variation of intensity. These three processes can be clearly noted for the A$_1^{'}$(1) modes apparent in the RS spectra of the GaSe and InSe crystals. In different materials, including S-TMDs~\cite{Late2014,Park2015,Yang2017,Gasanly2002}, one finds that both the energy and the linewidth of phonon mode vary with temperature. This temperature dependence was attributed to the anharmonic terms in the vibrational potential energy~\cite{Balkanski1983}. As we investigate the temperature effect on excitation conditions of RS, we focus our attention on its influence on phonon-modes intensities. While in almost the whole range of the used temperature to about 300 K, there is no significant change of peaks intensities, a strong quenching of the RS signal for all the observed lines in GaSe and InSe is apparent at the highest temperature of 350~K. This kind of behaviour can be ascribed to the variation of the excitation conditions of RS. As we measured RS spectra with the fixed laser energy for a given material, the temperature affects the band structure of the studied GaSe and InSe resulting in the change of the relative energy separation between laser energy and the optical band-gap energy for GaSe or B-transition energy for InSe. Interestingly, the vanishing of phonon modes at $T$=350~K for GaSe is accompanied with the emergence of the broad emission covering the whole range of the detected energies, which can be ascribed to the emission in the vicinity of the optical band gap of GaSe. For the InSe, a similar quenching of phonon modes at $T$=350~K is followed by the emission extending from laser liner in both Stokes and anti-Stokes scatterings, which may be attributed to the resonant excitation of the B transition.

\begin{figure}[t]
	\includegraphics[width=0.5\textwidth]{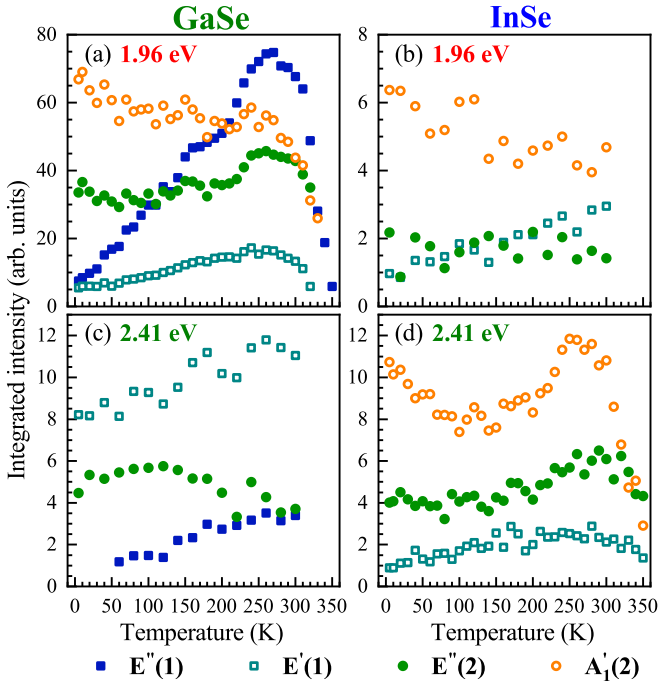}
	\caption{Temperature dependence of the phonon-modes intensities of (a),(c) GaSe and (b),(d) InSe under excitation of (a),(b) 1.96 eV and (c),(d) 2.41 eV.} \label{fig:3}
\end{figure}

\begin{figure}[b]
	\includegraphics[width=0.5\textwidth]{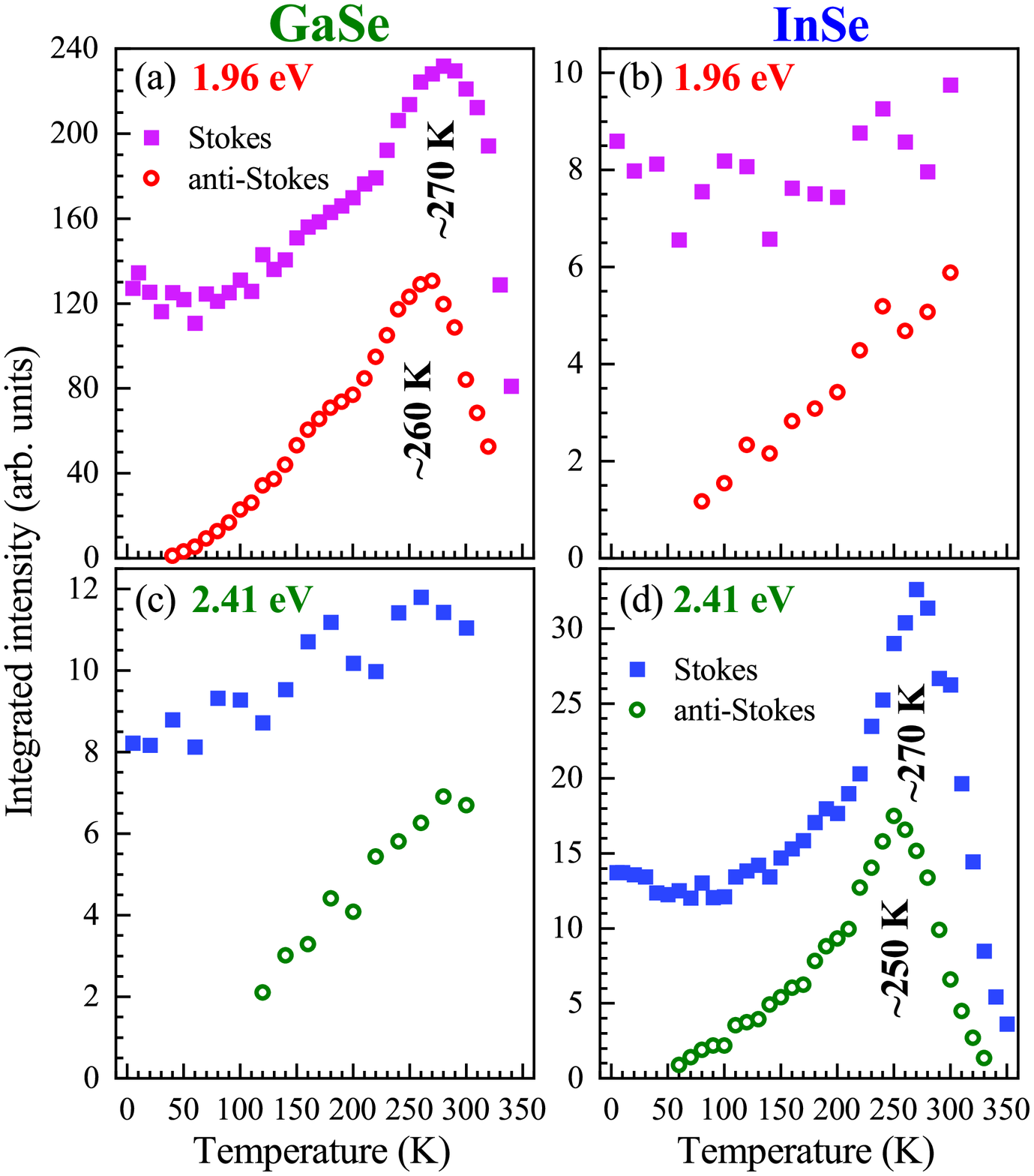}
	\caption{Temperature evolution of the Stokes- (full points) and anti-Stokes-related (open points) intensities for the A$_1^{'}$(1) modes measured on (a),(c) GaSe and (b),(d) InSe under excitation of (a),(b) 1.96 eV and (c),(d) 2.41 eV.} \label{fig:4}
\end{figure}

To investigate in more detail the temperature effect on the phonon-modes intensities, we fitted the observed phonon modes with Gaussian functions. The obtained intensity evolutions of the RS peaks under both excitation energies (1.96 eV and 2.41 eV) as a function of temperature are presented in Figs.~\ref{fig:3} and \ref{fig:4}. First, we can focus on the temperature evolutions of the phonon modes shown in Fig.~\ref{fig:3}. It is observed that their intensity dependencies for both materials are very similar, if the results are compared diagonally, $i.e.$ Figs.~\ref{fig:3}(a) and \ref{fig:3}(c); Figs.~\ref{fig:3}(b) and \ref{fig:3}(c). For the former case, $i.e.$ Figs.~\ref{fig:3}(a) and \ref{fig:3}(c), a small temperature effect on the modes intensities is apparent up to around 270~K, while at higher temperatures their intensities are quickly vanished up to almost zero. The most spectacular results are observed for the E$^{''}$(1) mode in GaSe and for the A$_1^{'}$(2) mode in InSe under excitation of 1.96~eV and 2.41~eV, respectively. Both these evolutions are characterized by well-resolved maxima located at about 270~K. In the latter case, $i.e.$ Figs.~\ref{fig:3}(b) and \ref{fig:3}(c), the corresponding dependences reveals monotonic behaviours (increase or decrease) as a function of temperature characterized by small slopes. These results suggest that the electron-phonon coupling for individual phonon modes is different, which can be understood by considering their symmetries with respect to the symmetries of the orbitals associated with the involved transitions, as it was reported for MoS$_2$~\cite{carvalho2015}. It is important to mention that similar effects, $i.e.$ the intensity maxima as a function of temperature, has not been reported so far for other LMs. Particularly, for thin layers of S-TMDs, only a growth or a decrease of the RS signal was observed in a broad temperature range from 5~K to 300~K~\cite{GolasaNano,Grzeszczyk2018,Molas2019}. The most interesting result is the strong vanishing of all modes intensities at temperatures above around 280 K under excitations of 1.96~eV and 2.41~eV in GaSe and InSe, respectively.

\subsection{Intensity profiles of the A$_1^{'}$(1) phonon modes \label{results:profiles}}

As we found that the most pronounced and temperature sensitive peak for both the materials is the A$_1^{'}$(1) mode, we decided to focus our attention only on its analysis. Figs.~\ref{fig:4}(a) and \ref{fig:4}(b) show the temperatures evolutions of the Stokes- and  anti-Stokes-related intensities of the A$_1^{'}$(1) modes. Particularly, analogous effects of temperature on phonon intensities shown in Fig.~\ref{fig:3} (comparison of results diagonally) are also observed for the A$_1^{'}$(1) in Fig.~\ref{fig:4}. For Figs.~\ref{fig:4}(a) and \ref{fig:4}(d), its well-resolved maxima located at about 250-270~K and its strong quenching in higher temperatures are apparent, while only monotonic evolutions are visible in Figs.~\ref{fig:4}(b) and \ref{fig:4}(c). It is interesting that the temperature-dependent A$_1^{'}$(1) intensity profiles for both GaSe and InSe are very similar, which may suggest analogous effects responsible for them. Moreover, it reflects the reported enhancement profile for phonon modes as a function of excitation energy in InSe in the vicinity of the B transition~\cite{Kuroda1978}. This suggests that the temperature variation under the fixed excitation energy can be treated as an analogous of the change of excitation energy. This leads us to the conclusion that resonant conditions of RS can be achieved under excitation of 1.96~eV for GaSe and 2.41~eV for InSe due to the energy proximity of the optical band gap and the B transition, respectively. Moreover, the observed small difference in the maxima positions between the Stokes- and anti-Stokes-related signals of about 10~K - 20~K may suggest that they are associated with the outgoing resonance~\cite{Vina1996}. This type of resonance is satisfied when the difference between the energies of excitation and the scattered (emitted) light is equal to the energy of phonon mode. In our case, we investigate the A$_1^{'}$(1) intensities in both the Stokes and anti-Stokes bands, which are related to its emission and absorption, respectively. Consequently, the difference in maxima positions should equal doubled energy of the  A$_1^{'}$(1).

\begin{figure}[t]
	\includegraphics[width=0.5\textwidth]{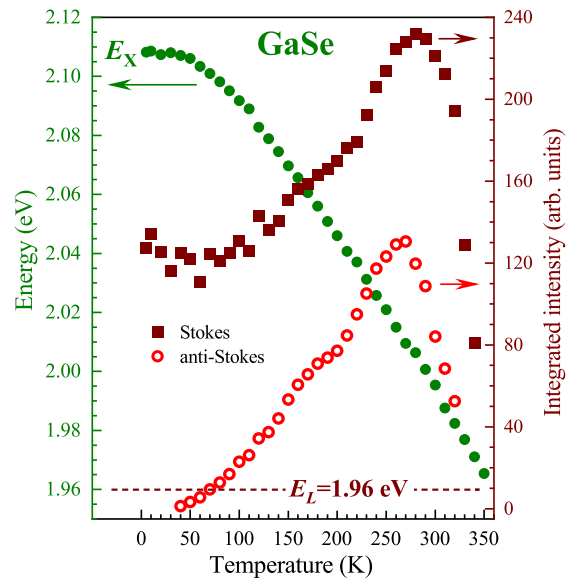}
	\caption{ The left hand scale refers to (green full points) the temperature evolution of the excitonic emission energy ($E_\textrm{X}$) in GaSe. The right hand scale presents the corresponding (red full points) Stokes- and (red open points) anti-Stokes-related intensities of the A$_1^{'}$(1) modes under the 1.96~eV excitation marked by dark red dashed horizontal line.}\label{fig:5}
\end{figure}

As the observed resonant conditions of RS in GaSe are due to the vicinity of the optical band gap, we measured its PL spectra in the whole range of temperatures. Fig. ~\ref{fig:5} presents a comparison of the temperature evolutions of the excitonic emission energy ($E_\textrm{X}$) in GaSe and the Stokes- and (red open points) anti-Stokes-related intensities of the A$_1^{'}$(1) modes under the 1.96~eV excitation. It is seen in the Figure that the maximum intensity of the A$_1^{'}$(1) observed at a temperature of about 270~K corresponds to the $E_\textrm{X}$ energy of $\sim$2.01~eV, which is much higher than the used laser energy of 1.96~eV, while the aforementioned vanishing of the mode intensity is apparent at the temperature of 350~K for which the excitation and emission energies are almost equal. The similar antiresonant behaviour was reported for the A$_1^{'}$(1) mode in GaSe at a temperature of about 80~K using a tunable laser~\cite{Hoff1974}. The reported minimum A$_1^{'}$(1) intensity is apparent when the excitation energy is around 50~meV below the exciton energy. In our opinion, this difference in the antiresonance energies can be described in terms of both the temperature tunablity of the band gap and of the electron-phonon coupling. The calculated band structures for $\varepsilon$-GaSe show a nearly direct band gap at the $\Gamma$ point of the Brillouin zone (BZ)~\cite{Do2015}. However, the detailed analysis of the GaSe band structure demonstrated in Ref.~\citenum{Do2015} reveals that the valence band maximum (VBM) moves away from $\Gamma$ towards the K-points, and the valence band takes the form of an inverted ''Mexican hat'' with the energy separation between the VBM and the $\Gamma$ point on the order of 8 meV. It may suggest that the measured emission energy may not correspond to the excitonic transition, which can qualitatively describe the difference between our results and those reported in Ref.\citenum{Hoff1974}. The origin of the antiresonant behaviour can be described in terms of quantum interference as it was reported for thin layers of MoTe$_2$~\cite{Miranda2017}, $i.e.$ the contributions to the Raman susceptibility from different regions (individual k-points) of the BZ add with particular signs (plus or minus). This may lead to enhancement or quenching of phonon-modes observed in the RS spectra. As our resonant conditions in GaSe occur in the vicinity of the optical band gap, it suggests that specific k-point close to the $\Gamma$ point of the BZ may give contributions of different sign resulting in vanishing of Raman signal. Moreover, we should also take into account that the electron-phonon interaction may lead to an increase of both the RS and PL signals, as it has been reported for monolayer WS$_2$~\cite{molasSR}. Consequently, the resonant conditions of Raman scattering may result in a strong enhancement of the observed emission (see Fig.~\ref{fig:2}, which triggers the vanishing of the Raman signal.

\begin{figure}[t]
	\includegraphics[width=0.5\textwidth]{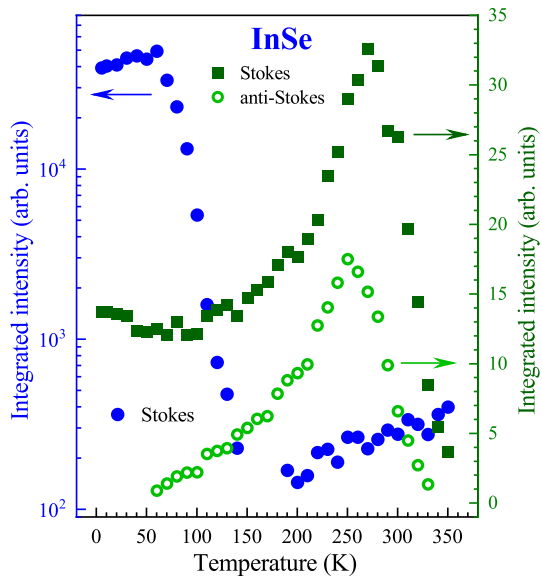}
	\caption{The left and right hand scales refers the temperature evolutions of the (blue and green full points) Stokes- and (green open points) anti-Stokes-related intensities of the A$_1^{'}$(1) modes under the 2.41~eV (green points) and 2.54~eV (blue points) excitations.}\label{fig:6}
\end{figure}

In the case of InSe, the description of the obtained results is more complex as we cannot measure directly the emission related to the B transition. Fig.~\ref{fig:6} presents the Stokes-related intensities for the A$_1^{'}$(1) mode under excitation of 2.54~eV and the corresponding Stokes and anti-Stokes intensities obtained with 2.41~eV excitation. We can appreciate that the maximum intensity of the A$_1^{'}$(1) mode apparent at a temperature of about 50~K under 2.54~eV excitation is significantly shifted as compared to $T$=270~K under 2.41~eV excitation. However, the most interesting is the antiresonant behaviour under 2.54~eV excitation, which reveals a quick decrease of the A$_1^{'}$(1) intensity of almost 2 orders of magnitude in the temperature range from 80~K to 140~K. It results in the complete vanishing of the A$_1^{'}$(1) mode for range from 150~K to 180~K. At higher temperatures, slow linear growth of the  A$_1^{'}$(1) intensity is visible. This shape of the A$_1^{'}$(1) intensity reflects the reported enhancement profile for this mode as a function of excitation energy in InSe in the vicinity of the B transition~\cite{Kuroda1978,Zolfaghari2019}. Consequently, we can speculate that the B transition energy varies from about 2.54~eV to 2.41~eV with temperature changes from about 50~K to 270~K. The origin of the antiresonance can be described similarly to GaSe in terms of quantum interference. However, as the used excitations are much larger than the InSe optical band gap, the possible contributions with different signs can be also taken with different points of BZ ($e.g.$ K or M points). It is important to mention that the analogous enhancement of the Raman scattering modes under different excitations in the vicinity of the B transitions was observed for thin layers of InSe~\cite{Molas2020}.

\section{Conclusions}
A comprehensive investigation of the temperature effect on the Raman scattering efficiency in $\varepsilon$-GaSe and $\gamma$-InSe crystals was reported. We found that the intensity of some phonon modes exhibits a strong variation as a function of temperature under excitation with specific energy due to the resonant conditions of RS. Moreover, a significant antiresonance behaviour accompanies the resonances at higher temperature, which leads to the vanishing of the modes intensities. The observed effects are discussed in terms of electron-phonon coupling and quantum interference effect.

\section*{Acknowledgements}
The work has been supported by the National Science Centre, Poland (grants no. 2017/24/C/ST3/00119 and 2017/27/B/ST3/00205).

\bibliographystyle{apsrev4-1}
\bibliography{biblio_raman}

\end{document}